\documentclass[11pt,a4]{article}

\usepackage{array,amsmath,amssymb,bm,color}
\usepackage[normalem]{ulem}
\newcommand{\F}{{\mathbf F}} 
\newcommand{\R}{{\mathbf R}} 
\newcommand{\C}{{\mathbf C}} 
\newcommand{\B}{{\mathbf B}}
\newcommand{\I}{{\mathbf I}}
\newcommand{\V}{{\mathbf V}} 
\newcommand{\U}{{\mathbf U}}

\newcommand{\T}{{\rm T }}
\newcommand{\tr}{{\rm tr\,}}

\newcommand{\be}{\begin{equation}}
\newcommand{\ee}{\end{equation}}
\newcommand{\ba}{\begin{array}}
\newcommand{\ea}{\end{array}}
\newcommand{\bqa}{\begin{eqnarray}}
\newcommand{\eqa}{\end{eqnarray}}
\newcommand{\bp}{\begin{pmatrix}}
\newcommand{\ep}{\end{pmatrix}}
\newcommand{\bv}{\begin{vmatrix}}
\newcommand{\ev}{\end{vmatrix}}

\title{${\bf U} = \mathbf{C}^{1/2}$ and its invariants in terms of  $\bf C$ and its invariants}
\author{N. H. Scott\thanks{Email: n.scott@uea.ac.uk},
School of Mathematics, University of East Anglia,\\ 
Norwich Research Park, Norwich NR4 7TJ, UK\\
}

\date{\today} 

\begin{document}
\maketitle

\begin{abstract}
We consider  $N\times N$ tensors for $N= 3,4,5,6$.  In the case $N=3$,
it is desired to find the three principal invariants $i_1, i_2, i_3$ of  $\U$ in terms of the three principal invariants $I_1, I_2, I_3$ of $\C=\U^2$.  Equations connecting the $i_\alpha$ and $I_\alpha$ are obtained by taking determinants of the factorisation 
\[\lambda^2\I-\C = (\lambda\I-\U)(\lambda\I+\U)\]
and comparing coefficients.  On eliminating $i_2$ we obtain a quartic equation with coefficients depending solely on  the  $I_\alpha$
 whose largest root is $i_1$.   Similarly, we may obtain a quartic equation whose largest root is $i_2$.   For $N=4$  we find that $i_2$ is once again the largest root of a quartic equation and so all the $i_\alpha$ are expressed in terms of the $I_\alpha$.  Then $\U$ and $\U^{-1}$ are expressed  solely in terms of $\C$, as for $N=3$.  For $N= 5$  we  find, but do not exhibit, a twentieth degree polynomial of which $i_1$ is the largest root and which has four spurious zeros.  We are unable to express  the $i_\alpha$ in terms of the $I_\alpha$ for $N=5$.  Nevertheless, $\U$ and $\U^{-1}$ are expressed in terms of powers of $\C$ with coefficients now depending on the  $i_\alpha$.   For $N=6$  we find, but do not exhibit, a 32 degree polynomial which has largest root $i_1^2$.  Sixteen of these roots are relevant but the other 16, which we exhibit, are spurious.  $\U$ and $\U^{-1}$ are expressed in terms of powers of $\C$.  The cases $N>6$ are  discussed.\\
 
\end{abstract}

\section{Introduction} 
\label{sec:1}

All tensors occurring in this paper are square of dimension $N\times N$.  In sections \ref{sec:1}--\ref{sec:3} we consider only the case $N=3$ except at the end of this section we briefly discuss the case $N=2$.  In sections \ref{sec:4}--\ref{sec:6} we discuss the cases $N=4$--6, respectively.

In terms of the deformation gradient $\F$ the right and left  Cauchy-Green strain tensors are defined by the symmetric positive definite tensors 
$ \C=\F^\T\F, \; \B=\F\F^\T, $
respectively, where $^\T$ indicates the transpose.   The polar decomposition theorem states that 
\[ \F=\R\U=\V\R, \quad \R\R^\T=\R^\T\R=\I, \]
in which $\R$ is a proper orthogonal, or rotation, tensor, $\U$ and $\V$ are respectively the right and left  stretch tensors and $\I$ is the identity tensor.  We see that
\( \C=\U^2, \)
so that
\[ \U=\C^{1/2}, \]
where $\U$ is the unique symmetric positive definite tensor square root of $\C$.

Assuming $\F$ is known then $\C$ is easy to calculate but $\U$ less so as it is a square root.  Denote the (necessarily positive) eigenvalues of $\C$ by $\lambda_i^2$ with corresponding eigenvectors $\bm{e}_i$ so that the eigenvalues of $\U$ are $\lambda_i$ with the same eigenvectors.  The $\lambda_i$ are the principal stretches. From the spectral representations
\[ \C= \sum_{i=1}^3 \lambda_i^2 \bm{e}_i \otimes \bm{e}_i, \quad
 \U= \sum_{i=1}^3 \lambda_i \bm{e}_i \otimes \bm{e}_i \]
 we see that one method would be to find the eigenvalues and eigenvectors of $\C$ numerically, square root the eigenvalues, and then use the second spectral representation above to determine $\U$.  However, this 
 does not result in a formula for $\U$.  
  Luehr \& Rubin \cite{luehr}  give a formula for the dyadic products $\bm{e}_i \otimes \bm{e}_i$ which for distinct $\lambda_i^2$, in the case $i=1$, reduces to
 \[ \bm{e}_1 \otimes \bm{e}_1 = 
 (\lambda_2^2 - \lambda_1^2)^{-1} (\lambda_3^2 - \lambda_1^2)^{-1} 
  (\C- \lambda_2^2\I)   (\C- \lambda_3^2\I),  \] 
 and similarly for $i=2,3$.   Thus we may express these dyadic products in terms of $\C$ and its eigenvalues without first calculating the eigenvectors  and so find a formula for $\U$.  
 %
 Jog \cite{jog} gives a different method of  determining dyadic products, one involving the inversion of Vandermonde matrices.  
Luehr \& Rubin's and Jog's  formulas express $\U$ and $\C$ in terms of the principal stretches $\lambda_i$ rather that the principal tensor invariants of $\U$ or $\C$.  Here we seek formulas of the latter type.


 Hoger \& Carlson \cite{hoger1} observed that one could use the Cayley-Hamilton theorem for $\U$ to determine $\U$ as a sum of powers of $\C$ without first needing to find the eigenvalues and eigenvectors of $\C$.  They did the same for $\U^{-1}$.  These ideas were further developed by Ting \cite{ting} who gave a formula for $\U^{-1}$ simpler than that of Hoger \& Carlson \cite{hoger1}.  See also Carroll \cite{carroll}.  One drawback of this method is that it produces coefficients of powers of $\C$ depending on the principal invariants of $\U$, rather than those of $\C$.   However, we can  find expressions for the invariants of $\U$ directly in terms of those of $\C$.
 
 Hoger \& Carlson \cite{hoger1} seek to determine the invariants of $\U$ in terms of those of $\C$ by finding a quartic equation satisfied by the first principal invariant of $\U$.  We extend these ideas in the present paper.  The tensors discussed so far have dimension $N=3$.  In addition to the case $N=3$ we shall discuss also the cases $N=4$,  $N=5$ and $N=6$.  Hoger \& Carlson \cite{hoger1} and Ting \cite{ting} also briefly discuss higher dimensional cases.

In his study of nonlinear fluid-structure interactions in flapping wing systems, Fitzgerald \cite[pages 67, 68]{fitzgerald} was compelled to write the invariants of $\U$ in terms of those of $\C$ in order to utilise a particular computer code.  He did this using the methods and notation of Norris \cite{norris1}, who shows how to express the principal invariants of $\U$ in terms of those of $\C$ in a symmetric manner using a single function call.  However, Fitzgerald \cite{fitzgerald}  observed that there were numerical problems in differentiating the functions of Norris close to the linear state.

This paper is constructed as follows.  In section \ref{sec:2} with $N=3$ we find the three squared principal stretches $\lambda_i^2$ in terms of the three principal invariants of $\C$  by solving the cubic characteristic equation.  Then in section \ref{sec:3} we see how to express the invariants of $\U$ in terms of those of $\C$, still for $N=3$, by means of the quartic equation of  Hoger \& Carlson \cite[after Eqn. (5.4)]{hoger1}.  We give a complete solution of this equation.  Moving to the case $N=4$ in section  \ref{sec:4}, we obtain a quartic equation with coefficients depending solely on the invariants of $\C$ for the second principal invariant $i_2$ of $\U$,  giving a complete solution for $i_2$ in terms of the roots of the resolvent cubic equation.  This method appears to be easier than solving the quartic characteristic equation directly.  In section  \ref{sec:5} we consider the case $N=5$.  Of course, there is no formula for the general solution of the quintic characteristic equation.  Instead, we proceed by seeking an equation for the first principal invariant $i_1$ of $\U$.  It is a polynomial equation of the twentieth degree and we can determine all the roots, the largest giving the value of $i_1$.
 In section  \ref{sec:6} we consider the case $N=6$.  We find a 32 degree polynomial in $w=i_1^2$, the largest root giving the value of $i_1^2$, and are able to find explicitly the 16 spurious roots.
We discuss the extension to $N>6$ in section  \ref{sec:7}.

For completeness we first dispose of the $N=2$ case, previously dealt with by Hoger \& Carlson \cite{hoger1} and Ting \cite{ting}.  With $I_1=\tr \C$, $I_2=\det \C$ and $\I$ denoting the unit tensor, we have
\[ \label{1.1a}
\U =\frac{1}{\sqrt{I_1+2\sqrt{I_2}}} \Big(\sqrt{I_2}\I+\C\Big),\quad
\U^{-1} = \frac{1}{\sqrt{I_2}\sqrt{I_1+2\sqrt{I_2}}} \Big((I_1+\sqrt{I_2})\I -\C\Big),
\]
in terms only of $\C$ and its invariants.  We do the same for $N=3,4$ but the coefficients are very much more complicated than for $N=2$.  For $N=5,6$ we find that we must also involve the invariants of $\U$ in expressions for $\U$ and $\U^{-1}$ in powers of $\C$.

\section{The squared  stretches in terms of the invariants of $\C$ for $N=3$} 
\label{sec:2}
\setcounter{equation}{0}

The eigenvalues of  $\C=\F^\T\F$ are the squared stretches $\lambda_1^2, \lambda_2^2, \lambda_3^2$ and we may assume the  ordering
\be \label{2.1a}
\lambda_1 \geq \lambda_2 \geq \lambda_3. 
\ee
The three principal invariants of $\C$ are defined by
\[ I_1 = \tr\C,\quad I_2 = \frac12\left[I_1^2 - \tr(\C^2) \right],\quad I_3 = \det\C, \]
which, in terms of the squared stretches, become
\be \label{2.2a}
I_1 = \lambda_1^2 + \lambda_2^2 + \lambda_3^2,\quad
I_2 =  \lambda_1^2\lambda_2^2 + \lambda_2^2\lambda_3^2 + \lambda_3^2\lambda_1^2,\quad
I_3 = \lambda_1^2\lambda_2^2\lambda_3^2.
\ee
The characteristic equation for $\C$ is
\be\label{2.3a}
\det(\lambda^2 \I  - \C) = 0, 
\ee
which on expansion, and writing $x=\lambda^2$, becomes
\be \label{2.4a}
x^3-I_1x^2+I_2x-I_3=0,
\ee
the roots of which  are the squared stretches $\lambda_1^2, \lambda_2^2, \lambda_3^2$.

By means of the substitution
\[ x=\tfrac13 I_1 + y\]
the cubic equation (\ref{2.4a}) is reduced to the standard form
\be \label{2.5a}
y^3=py+q
\ee
in which
\be \label{2.6a}
\begin{aligned}
p &=\tfrac13 (I_1^2-3I_2),\\
q &=\tfrac{1}{27} (2I_1^3-9I_!I_2+27I_3 ).
\end{aligned}
\ee
From the definitions (\ref{2.2a}) we can show that
\be \label{2.7a}
p = \tfrac16[( \lambda_1^2-\lambda_2^2)^2  + ( \lambda_2^2-\lambda_3^2)^2 +( \lambda_3^2-\lambda_1^2)^2] \geq0,
\ee
vanishing only when $\lambda_1 = \lambda_2 = \lambda_3$.  A result seemingly new to the literature is the factorisation
\be\label{2.8a}
q = \tfrac{1}{27} 
( 2\lambda_1^2-\lambda_2^2-\lambda_3^2)
( 2\lambda_2^2-\lambda_3^2-\lambda_1^2)
( 2\lambda_3^2-\lambda_1^2-\lambda_2^2),
\ee
with the ordering (\ref{2.1a}) forcing the first bracket to be positive, the last to be negative, and the middle one to have either sign, or to vanish (if $\lambda_2^2 = \tfrac12(\lambda_1^2+\lambda_3^2)$).  To summarise, $p$ vanishes if and only if the $\lambda_i^2$ are all equal and $q$ vanishes if and only if the $\lambda_i^2$ are in arithmetic progression (possibly with zero common difference).

Cardano's solution for the cubic equation  (\ref{2.5a}) gives the roots
\be\label{2.9a}
y = \left(\frac{q}{2} + \sqrt{D}\right)^{1/3} +
\left(\frac{q}{2} - \sqrt{D}\right)^{1/3},
\ee
where the discriminant $D$ is given by
\be\label{2.10a}
\begin{aligned}
D &=\left(\frac{q}{2}\right)^2 - \left(\frac{p}{3}\right)^3\\
  &= -\frac{1}{108}\left[I_1^2I_2^2+18I_1I_2I_3 - 4I_1^3I_3-4I_2^3-27I_3^2\right]\\
    &=  - \frac{1}{108} ( \lambda_1^2-\lambda_2^2)^2  ( \lambda_2^2-\lambda_3^2)^2 ( \lambda_3^2-\lambda_1^2)^2\\
    &\leq 0,
  \end{aligned}
\ee
vanishing only if at least two of the $\lambda_i$ are equal.  Franca \cite[Eqns (34), (38)]{franca} also gives expressions equivalent to (\ref{2.7a}) and (\ref{2.10a}).  We denote the argument of the complex number  $\frac{q}{2} + \sqrt{D}$ by $3\theta$ and note that its modulus is 
\[\left| \frac{q}{2} + \sqrt{D}\right| = \left|\frac{q}{2} + i\sqrt{|D|} \right| = \left(\frac{p}{3}\right)^{3/2}, \]
so that $\theta$ may be determined from
\[\cos 3\theta = \frac{q/2}{(p/3)^{3/2}}.   \]
On defining the angles
\be \label{2.11a}
\theta_1 = \frac13 \cos^{-1}\left( \frac{q/2}{(p/3)^{3/2}}\right),\quad \theta_2 = \theta_1- \frac{2\pi}{3},\quad \theta_3 = \theta_1 +  \frac{2\pi}{3}, 
\ee
choosing the branch of $\cos^{-1}$ with $0\leq\theta_1\leq\pi$, we find that the squared  stretches, namely, the three roots of the characteristic equation (\ref{2.4a}), are given by
\be\label{2.12a}
\lambda_i^2 = \frac13 I_1 + 2\left(\frac{p}{3}\right)^{1/2} \cos\theta_i,\quad i=1, 2, 3.
\ee
The angles $\theta_i$ have been chosen so that the roots (\ref{2.12a}) satisfy the ordering (\ref{2.1a}).  Equations (\ref{2.12a}) express the $\lambda_i^2$ purely in terms of the invariants of $\C$ by means of (\ref{2.6a}) and (\ref{2.11a}).

The roots (\ref{2.12a}) could have been obtained using the identity
\[ 4\cos^3\theta - 3\cos\theta \equiv \cos3\theta \]
instead of Cardano's formula.

\section{The invariants of $\U$ expressed in terms of those of $\C$ for  $N=3$} 
\label{sec:3}
\setcounter{equation}{0}

The invariants of $\U=\C^{1/2}$ are defined in terms of the principal stretches $\lambda_i$ by 
\be \label{3.1a}
i_1 = \lambda_1 + \lambda_2 + \lambda_3,\quad
i_2=  \lambda_1\lambda_2 + \lambda_2\lambda_3 + \lambda_3\lambda_1,\quad
i_3 = \lambda_1\lambda_2\lambda_3
\ee
and we seek to determine them in terms of the invariants of $\C$.

In terms of the invariants of $\C$ we see from (\ref{2.12a}) that
\be\label{3.2a}
i_1=\lambda_1+\lambda_2+\lambda_3,\mbox{  where  } 
\lambda_i = \sqrt{\frac13 I_1 + 2\left(\frac{p}{3}\right)^{1/2} \cos\theta_i},\quad i=1, 2, 3,
\ee
with $p$ given by (\ref{2.6a})$_1$ and $\theta_i$ given by (\ref{2.11a}).  It is possible to write $i_1$ defined at (\ref{3.1a})$_1$ in terms of only one of the principal stretches, say $\lambda_1$ given by (\ref{3.2a}) with $i=1$, and the invariants of $\C$, by noting that
\[ \lambda_2+\lambda_3 = \sqrt{\lambda_2^2+\lambda_3^2   + 2\lambda_2\lambda_3 } = \sqrt{I_1-\lambda_1^2+2\sqrt{I_3}/\lambda_1} \]
so that from (\ref{3.2a})
\be\label{3.3a}
i_1 = \lambda_1 + \sqrt{I_1-\lambda_1^2+2\sqrt{I_3}/\lambda_1}.
\ee
This is the expression for $i_1$ derived by Hoger \& Carlson \cite[Eqn. (5.5)]{hoger1}, though in a different notation and by a different method.  The second line of their expression is redundant, as noted also by Norris \cite{norris1}. 

An alternative method of determining $i_1$ is to seek a single equation for it following the method of Hoger \& Carlson \cite{hoger1}.   On replacing $x$ by $\lambda^2$ in the characteristic equation (\ref{2.3a}) we observe the factorisation
\[
\begin{aligned}
\lambda^2\I  -\C
&= (\lambda\I  -\U) (\lambda\I  +\U) \mbox{\quad  leading to } \\
 \det(\lambda^2\I  -\C) 
&= \det(\lambda\I  -\U) \det(\lambda\I  +\U) \\
&= (\lambda^3-i_1\lambda^2+i_2\lambda-i_3)
       (\lambda^3+i_1\lambda^2+i_2\lambda+i_3)
\end{aligned}
\]
involving the invariants of $\U$.  On multiplying out these parentheses and comparing with (\ref{2.4a}) we find that
\be\label{3.4a}
 i_1^2 =I_1+2i_2,\quad
 i_2^2 =I_2+2i_1i_3,\quad
 i_3^2 =I_3.
\ee
On eliminating $i_2$ and  $i_3$ we see that $y=i_1$ is a root of the quartic equation
\be\label{3.5a}
y^4   -2I_2y^2 -8\sqrt{I_3}y+  I_1^2-4I_2 =0.
\ee
Hoger \& Carlson \cite[after Eqn. (5.4)]{hoger1}  also obtained equation (\ref{3.5a})  for $i_1$. 

We use Ferrari's method to solve the general reduced quartic equation
\be\label{3.6a}
y^4 + py^2 + qy+ r =0,
\ee
with arbitrary coefficients $p$, $q$ and $r$, by first rewriting it in the form
\be \label{3.7a}
\begin{aligned}
\left(y^2+\frac{p}{2}+2n\right)^2 &= 4 \left(\sqrt{n} y - \frac{q/8}{\sqrt{n}}\right)^2 \\
&\quad \mbox{}+  \frac{4}{n} \left\{ n^3 + \left(\frac{p}{2}\right) n^2 + \left( \left(\frac{p}{2}\right)^2 - r\right)\frac{n}{4} - \left(\frac{q}{8}\right)^2 \right\},
 \end{aligned}
\ee
equivalent to (\ref{3.6a}) for any non-zero choice of $n$.  Thus, if $n$ is any root of the resolvent cubic equation
\be \label {3.8a}
n^3 + \left(\frac{p}{2}\right) n^2 + \left( \left(\frac{p}{2}\right)^2 - r\right)\frac{n}{4} - \left(\frac{q}{8}\right)^2 = 0,
\ee
equation  (\ref{3.7a}) can be square rooted to give two quadratic equations in $y$ and so all four roots of (\ref{3.6a}) can be determined explicitly.    With $n_i$, $i=1,2,3$, denoting the three roots of (\ref{3.8a}), we see from  (\ref{3.8a}) that
\[
\frac{p}{2} = -(n_1+n_2+n_3),\quad
\frac{q}{8} = \sqrt{n_1}\sqrt{n_2}\sqrt{n_3},
\]
taking positive square roots,
and then are able to deduce that the four roots of (\ref{3.7a}),  and therefore of (\ref{3.6a}),  are
\be \label{3.9a}   \begin{aligned}
\sqrt{n_1} + \sqrt{n_2} + \sqrt{n_3},\\
\sqrt{n_1} - \sqrt{n_2} - \sqrt{n_3},\\
- \sqrt{n_1} + \sqrt{n_2} - \sqrt{n_3},\\
- \sqrt{n_1} - \sqrt{n_2} + \sqrt{n_3},
\end{aligned} 
\ee
These results are similar to those obtained by Euler's method of solving the quartic equation.

In the present case of the quartic equation (\ref{3.5a}) we find that the resolvent cubic equation (\ref{3.8a}) reduces to
\be \label{3.10a}
n^3-I_1n^2+I_2n-I_3=0,
\ee
identical to the characteristic equation (\ref{2.4a}) of $\C$.  Thus $n$ may be taken  equal to any one of the squared stretches $\lambda_i^2$, given in terms of the invariants of $\C$ by  (\ref{2.12a}).   Franca \cite{franca} also remarks on the close association of  the cubic equations (\ref{3.10a}) and (\ref{2.4a}) with the quartic equation (\ref{3.5a}).

We find that the four roots (\ref{3.9a})  of  (\ref{3.5a}) are
\be \label{3.11a} \begin{aligned}
y = \lambda_1 + \lambda_2 + \lambda_3,\\
 \lambda_1 - \lambda_2 - \lambda_3,\\
 -\lambda_1 + \lambda_2 - \lambda_3,\\
 -\lambda_1 - \lambda_2 + \lambda_3, 
 \end{aligned}
 \ee
rewritten more concisely as
\be\label{3.12a}
y = i_1,\quad 2\lambda_1-i_1,\quad 2\lambda_2-i_1,\quad 2\lambda_3-i_1,
\ee
in descending order of size because of (\ref{2.1a}).  The first root must be positive and the last two negative but the second may be positive, negative, or zero (if $\lambda_1=\lambda_2+\lambda_3$).  Then $i_1$ is given by the largest positive root of (\ref{3.5a}), namely (\ref{3.11a})$_1$,  the other three roots being regarded as spurious.  Thus, once again we see that $i_1$ is given by (\ref{3.3a}).  Equation (\ref{3.3a}) was derived directly from the cubic characteristic equation (\ref{2.4a}) whereas here it was derived from the quartic equation (\ref{3.5a}) for~$i_1$.

Hoger \& Carlson \cite{hoger1} claim incorrectly that (\ref{3.5a})  has a unique positive root and proceed on this assumption.  We have seen that there may be two positive roots and that the larger must be taken to be $i_1$.
Sawyers \cite{sawyers} gave an example to show that there may be more than one positive root and saw that one must take the larger to give $i_1$.  Sawyer's \cite{sawyers} example also exhibits two negative roots as proved above, though he did not remark on this.

Franca \cite[Eqn. (31)]{franca} and Norris \cite[Eqn. (12a)]{norris1} also give results equivalent to our (\ref{3.3a}) and Franca gives the four roots (\ref{3.11a}) as do Bouby et al. \cite{bouby}.  Franca remarks that it is computationally more efficient to calculate $i_1$ using just one of the principal stretches $\lambda_i$ rather than calculating all three.

Bouby et al. \cite{bouby} note that the principal invariants of $\C$ are invariant under sign change of any of the stretches $\lambda_i$. But we can go further than this. The quantity $\sqrt{I_3}$ occurring in equation (\ref{3.5a}) is really $i_3=\lambda_1\lambda_2\lambda_3$ and so is invariant only if \emph{two} of the $\lambda_i$ change sign.  Then equation (\ref{3.5a}) is invariant under two sign changes of the $\lambda_i$ and so therefore must be the set of its roots.  We know that (\ref{3.11a})$_1$ is one root and so the others, namely (\ref{3.11a})$_{2,3,4}$, can all be obtained by changing \emph{two} signs among the $\lambda_i$.

We turn now to the second invariant $i_2$.  On eliminating $i_1$ from (\ref{3.4a})  we find that $z= i_2$ is a root of the quartic equation
\be\label{3.13a}
z^4 -2I_2z^2-8I_3z+I_2^2-4I_1I_3 = 0,
\ee
which can be solved as was  (\ref{3.5a}) to obtain in place of  (\ref{3.12a})
\be\label{3.14a}
z = i_2,\quad 2\lambda_1\lambda_2-i_2,\quad 2\lambda_1\lambda_3-i_2,\quad 2\lambda_2\lambda_3-i_2,
\ee
again arranged in descending order because of (\ref{2.1a}). The first root is positive, the last two negative and the second of either sign (or zero if $\lambda_3^{-1}=\lambda_1^{-1}+\lambda_2^{-1}$).  These roots could also have been obtained directly by inserting the $i_1$ values from (\ref{3.12a})  into (\ref{3.4a})$_1$.

Alternatively,  arguing from its definition (\ref{3.3a})$_2$ and using the methods used to derive (\ref{3.3a}),  we see  that $i_2$ is given solely in terms of the invariants of $\C$ by
\be  \label{3.15a}
 i_2 = \sqrt{I_3}/\lambda_1 +  \lambda_1\sqrt{I_1-\lambda_1^2+2\sqrt{I_3}/\lambda_1},
\ee
since  $\lambda_1$ is so given by (\ref{2.12a}).  Norris \cite[Eqn. (12b)]{norris1} also gives this result.

\subsection{$\U$ and $\U^{-1}$ expressed in terms of $\C$ and its invariants}

For $N=3$ we shall need the quantity
\be \label{3.16a}
\begin{aligned}
\nu_3 
    &= i_1i_2-i_3  \\
    &= (\lambda_1+\lambda_2)(\lambda_2+\lambda_3)(\lambda_3+\lambda_1) \\
    &= (i_1-\lambda_1) (i_1-\lambda_2) (i_1-\lambda_3)  \\
    &= (i_1-\lambda_1)(\lambda_1i_1+ \sqrt{I_3}/\lambda_1).  
\end{aligned}
\ee
From the definitions (\ref{3.1a}) we see that $i_1i_2-i_3 $ vanishes on putting $\lambda_1=-\lambda_2$ leading to the factorisation  (\ref{3.16a})$_2$.  Further manipulation leads to (\ref{3.16a})$_4$ in which $\nu_3$ is
expressed solely in terms of the invariants of $\C$ as $\lambda_1$ is so expressed by (\ref{2.12a})  and $i_1$  by (\ref{3.3a}).

In order to express $\U$ in terms of $\C$ and its invariants for $N=3$ we follow the method of Ting \cite{ting} and multiply the Cayley-Hamilton theorem for $\U$  by $\U$ and eliminate $\U^3$ from the resulting two equations to obtain, after use of (\ref{3.4a})$_1$,
\be \label{3.17a} \nu_3\U = i_1\sqrt{I_3}\I+(I_1 + i_2)\C-\C^2  \ee
with $\nu_3$  given by (\ref{3.16a})$_4$, $i_1$  given by (\ref{3.3a}) and   $i_2$  given by (\ref{3.15a}), so that $\U$ is given in terms only of $\C$ and its invariants.  Hoger \& Carlson \cite[(3.7)]{hoger1} and  Ting \cite[(2.7)]{ting} give results equivalent to  (\ref{3.17a}).   
 Jog \cite[(13)]{jog} is also equivalent to (\ref{3.17a}).

To get $\U^{-1}$ we follow Ting \cite{ting} and multiply (\ref{3.17a}) by $\C^{-1}$ and then use the Cayley-Hamilton theorem for $\C$ to eliminate $\C^{-1}$.  Finally,
\be \label{3.18a}
\nu_3\U^{-1} = \left(I_1+i_2+\frac{i_1I_2}{\sqrt{I_3}}  \right) \I 
- \left( 1+ \frac{i_1I_1}{\sqrt{I_3}} \right)\C 
+ \frac{i_1}{\sqrt{I_3}}\C^2.
\ee
Hoger \& Carlson \cite[(4.2)]{hoger1} and  Ting \cite[(3.2)]{ting}  give results equivalent to  (\ref{3.18a}).
 Jog \cite[equation following (13)]{jog}  is also equivalent to (\ref{3.18a}).

Equations (\ref{3.17a}) and (\ref{3.18a})  express $\U$ and $\U^{-1}$, respectively,  in terms of $\C$ with equations  (\ref{3.3a}), (\ref{3.15a}) and  (\ref{2.12a}) giving $i_1$, $i_2$ and $\lambda_1$ in terms of $\C$.

\section{The invariants of $\U$ expressed in terms of those of $\C$ for $N=4$} 
\label{sec:4}
\setcounter{equation}{0}

We now  consider higher dimensional cases, first with $N=4$, so that $\C$ is now a $4\times4$ positive definite tensor.  As before, we seek the four principal invariants of $\U$ in terms of those of $\C$.

The four principal invariants of $\C$ are defined by
\[ 
\begin{aligned}
I_1 &= \tr\C,\quad I_2 = \frac12\left[I_1^2 - \tr(\C^2) \right], \\ 
I_3 &= \frac13\left[ \tr(\C^3)  - I_1^3  + 3I_1I_2 \right], \quad I_4 = \det\C,
\end{aligned}
 \]
which, in terms of the squared stretches, become
\be \label{4.1a}
\begin{aligned}
I_1 &= \lambda^2_1 + \lambda^2_2 + \lambda^2_3 + \lambda^2_4,\\
I_2 &=  \lambda^2_1\lambda^2_2  +  \lambda^2_1\lambda^2_3 +  \lambda^2_1\lambda^2_4 + \lambda^2_2\lambda^2_3 + \lambda^2_2\lambda^2_4 + \lambda^2_3\lambda^2_4,\\
I_3 &= \lambda^2_1\lambda^2_2\lambda^2_3 +  \lambda^2_1\lambda^2_2\lambda^2_4+  \lambda^2_1\lambda^2_3\lambda^2_4 +  \lambda^2_2\lambda^2_3\lambda^2_4,\\
I_4 &= \lambda^2_1\lambda^2_2\lambda^2_3\lambda^2_4.
\end{aligned}
\ee

The characteristic equation (\ref{2.4a}) for $\C$ becomes
\be \label{4.2a}
x^4-I_1x^3+I_2x^2-I_3x+I_4=0,
\ee
the roots of which  are the four squared stretches.

The invariants of $\U=\C^{1/2}$ are defined in terms of the principal stretches by 
\[
\begin{aligned}
i_1 &= \lambda_1 + \lambda_2 + \lambda_3 + \lambda_4,\\
i_2 &=  \lambda_1\lambda_2  +  \lambda_1\lambda_3 +  \lambda_1\lambda_4 + \lambda_2\lambda_3 + \lambda_2\lambda_4 + \lambda_3\lambda_4,\\
i_3 &= \lambda_1\lambda_2\lambda_3 +  \lambda_1\lambda_2\lambda_4+  \lambda_1\lambda_3\lambda_4 +  \lambda_2\lambda_3\lambda_4,\\
i_4 &= \lambda_1\lambda_2\lambda_3\lambda_4.
\end{aligned}
\]
The corresponding characteristic equation for $\U$ is
\[
x^4-i_1x^3+i_2x^2-i_3x+i_4=0,
\]
the roots of which  are the four stretches.

We seek to determine the invariants $i_\alpha$ of $\U$ in terms of the  invariants $I_\alpha$ of $\C$ and so derive
the following identities  in the same way that  (\ref{3.4a}) were derived:
\be \label{4.3a}
\begin{aligned}
i_1^2 &= I_1+2i_2,\\
i_2^2 &=  I_2+2i_1i_3-2i_4,\\
i_3^2 &= I_3+2i_2i_4,\\
i_4^2 &= I_4.
\end{aligned}
\ee
We could eliminate $i_2$ and $i_3$ between equations (\ref{4.3a}) and obtain an eighth degree equation in $i_1$ or we could eliminate $i_1$ and $i_2$ and obtain an eighth degree equation in  $i_3$.   However, we shall see that  if we eliminate instead $i_1$ and $i_3$ we shall obtain a quartic equation in $i_2$.
To this end
equation (\ref{4.3a})$_{2}$ may be written
\be \label{4.4a}
2i_1i_3 = i_2^2  -  I_2 + 2\sqrt{I_4}, 
\ee
which on squaring allows  $i_1^2$ and $i_3^2$ to be eliminated  in favour of $i_2$, using (\ref{4.3a})$_{1, 3}$, respectively,  to show that $y=i_2$ is one root of the quartic equation  (\ref{3.6a}) with
\be \label{4.5a}
\begin{aligned}
p &=  -2(I_2+6\sqrt{I_4}),\\
q &=  -8(I_1\sqrt{I_4} + I_3),\\
r &=   I_2^2-4I_1I_3 - 4I_2\sqrt{I_4} + 4 I_4.
\end{aligned}
\ee

With the coefficients  (\ref{4.5a})  the resolvent cubic equation (\ref{3.8a}) becomes
\be \label{4.6a}
n^3 - (I_2+6\sqrt{I_4})n^2 + (I_1I_3+4I_2\sqrt{I_4}+8I_4)n - (I_1\sqrt{I_4} + I_3)^2  = 0.
\ee
The identity
\[ I_2+6\sqrt{I_4} \equiv (\lambda_1\lambda_2+\lambda_3\lambda_4)^2 +  (\lambda_1\lambda_3+\lambda_2\lambda_4)^2 + (\lambda_1\lambda_4+\lambda_2\lambda_3)^2 \]
leads us to suspect that the three roots of (\ref{4.6a}) might be
\be \label{4.7a}
n_1 = (\lambda_1\lambda_2+\lambda_3\lambda_4)^2,\quad 
n_2 =  (\lambda_1\lambda_3+\lambda_2\lambda_4)^2,\quad  n_3 =  (\lambda_1\lambda_4+\lambda_2\lambda_3)^2 
\ee
and it may be verified by direct calculation that this is indeed the case.
Then
\be \label{4.8a}
\lambda_1 \geq \lambda_2 \geq \lambda_3 \geq \lambda_4
\implies n_1  \geq n_2 \geq n_3.
\ee

We see from  (\ref{4.6a}) that
\[
n_1+n_2+n_3 = I_2+6\sqrt{I_4}, \quad
\sqrt{n_1}\sqrt{n_2}\sqrt{n_3} = I_1\sqrt{I_4} + I_3,
\]
taking positive square roots of  (\ref{4.7a}),
and then are able to deduce that the four roots of (\ref{3.6a}), with coefficients (\ref{4.5a}), are given by (\ref{3.9a}) with $n_i$ now given by (\ref{4.7a}).
The first of these, the largest, is clearly $i_2$ and the other three are spurious roots.
In the same way that (\ref{3.3a}) was obtained we can show that 
\be \label{4.9a}
i_2 = \sqrt{n_1} + 
\sqrt{I_2 + 6\sqrt{I_4} - n_1 + 2(I_1\sqrt{I_4} + I_3)/\sqrt{n_1}},
\ee
expressing $i_2$ in terms of the invariants of $\C$ and only one of the roots of  (\ref{4.6a}), say~$n_1$. 

It remains to express $n_1$ itself in terms of  the invariants of $\C$  by obtaining an explicit solution of the cubic equation  (\ref{4.6a}) in the same way that the solutions  (\ref{2.12a}) were obtained for the characteristic equation  (\ref{2.4a}). 
Writing
\[ n=\tfrac13(I_2+6\sqrt{I_4}) + y  \]
in  (\ref{4.6a})  reduces it to the form (\ref{2.5a}) where now $p$ and $q$ are given by
\be \label{4.10a}
\begin{aligned}
p &=\tfrac13 (I_2^2-3I_1I_3+12I_4),\\[1mm]
q &=\tfrac{1}{27} (2I_2^3-9I_!I_2I_3+27I_1^2I_4+27I_3^2 -72I_2I_4),
\end{aligned}
\ee
in place of (\ref{2.6a}). 
Then the solution (\ref{2.12a})  is replaced by
\be\label{4.11a}
n_i = \frac13 (I_2+6\sqrt{I_4}) + 2\left(\frac{p}{3}\right)^{1/2} \cos\theta_i,\quad i=1, 2, 3,
\ee
the angles $\theta_i$ still defined by (\ref{2.11a})  and  the roots (\ref{4.11a}) satisfy the ordering (\ref{4.8a}).
On replacing $\lambda_i^2$  by $n_i$ in equations (\ref{2.7a}) and (\ref{2.8a}) they remain valid for $p$ and $q$ defined by (\ref{4.10a}).  

Inserting $n_1$ defined by (\ref{4.11a}) with $i=1$ into (\ref{4.9a}) now gives $i_2$ in terms only of the invariants of $\C$.  
Then $i_1$ is obtained  from (\ref{4.3a})$_1$ and $i_3$ from (\ref{4.3a})$_3$ by taking positive square roots, with $i_2$ continuing to be given by (\ref{4.9a}):
\be \label{4.12a}
i_1 = \sqrt{I_1+2i_2},\quad i_3 = \sqrt{I_3+2i_2\sqrt{I_4}}.
\ee
Thus for $N=4$ we have expressed the invariants of $\U$ entirely in terms of those of $\C$ by means of the equations (\ref{4.9a}), (\ref{4.11a}) with $i=1$, and (\ref{4.12a}).

Hoger \& Carlson \cite{hoger1} suggest obtaining the $\lambda_i$ by solving the quartic  characteristic equation  (\ref{4.2a}) algebraically but we find this more cumbersome than the above method.

\subsection{$\U$ and $\U^{-1}$ expressed in terms of $\C$ and its invariants}

For $N=4$ we need to define
\be \label{4.13a}
\nu_4 = i_1i_2i_3 - i_3^2 - i_1^2i_4.
\ee
By putting $\lambda_1 = -\lambda_2$ in (\ref{4.13a}) we see that   $\nu_4$ has the factor $\lambda_1+\lambda_2$ leading to the factorisation
\be \label{4.14a}
\nu_4 = (\lambda_1+\lambda_2) (\lambda_1+\lambda_3) 
(\lambda_1+\lambda_4) (\lambda_2+\lambda_3) 
(\lambda_2+\lambda_4) (\lambda_3+\lambda_4) .
\ee
We take the Cayley-Hamilton theorem for $\U$ and multiply it successively by $\U$ and $\U^2$.  From the resulting three equations we eliminate $\U^5$ and $\U^3$ in favour of $\U$ to obtain
\be\label{4.15a}\begin{aligned}
\nu_4\U &=  (i_1i_2-i_3)\sqrt{I_4}\,\I \\
 &\quad\mbox{}+(i_1i_2^2 - i_1^2i_3 - i_2i_3 + i_1\sqrt{I_4} )\C\\ 
 &\quad\mbox{} - (i_1^3 - 2i_1i_2 +i_3)\C^2 + i_1\C^3.
\end{aligned}
\ee
On multiplying by $\C^{-1}$ and arguing as before we obtain
\be \label{4.16a}
\begin{aligned}
\nu_4\U^{-1}  &= \left[ i_1i_2^2 - i_1^2i_3 - i_2i_3 + i_1\sqrt{I_4} + (i_1i_2-i_3)\frac{I_3}{\sqrt{I_4}}  \right]\I \\
&\quad\mbox{} - \left[ i_1^3 - 2i_1i_2 +i_3 + (i_1i_2-i_3)\frac{I_2}{\sqrt{I_4}}  \right]\C \\
  &\quad\mbox{} + \left[ i_1 + (i_1i_2-i_3)\frac{I_1}{\sqrt{I_4}}  \right] \C^2 - (i_1i_2-i_3)\frac{1}{\sqrt{I_4}} \C^3.
\end{aligned}
\ee
Equations (\ref{4.15a}) and (\ref{4.16a})  express $\U$ and $\U^{-1}$, respectively,  in terms of $\C$, using  (\ref{4.9a}), (\ref{4.11a}) and (\ref{4.12a}) to express the invariants of~$\U$ in terms of those of $\C$.


\section{An equation for $i_1$ in terms of the invariants of $\C$ for $N=5$} 
\label{sec:5}
\setcounter{equation}{0}

In the case $N=5$ the five principal invariants of $\C$ are defined by
\be \label{5.1a}
\begin{aligned}
I_1 &= 
\tr\C,\quad I_2 = \frac12\left[  I_1^2 - \tr(\C^2)  \right],\quad 
I_3 = \frac13\left[  \tr(\C^3)  - I_1^3  + 3I_1I_2 \right], \\
I_4  &= 
\frac14\left[ I_1^4-4I_1^2I_2+4I_1I_3 + 2I_2^2-\tr(\C^4)  \right], \quad I_5 = \det\C,
\end{aligned} 
\ee
which, in terms of the squared stretches, become
\be \label{5.2a}
\begin{aligned}
I_1 &= \lambda^2_1 + \lambda^2_2 + \lambda^2_3 + \lambda^2_4 + \lambda^2_5 ,\\
I_2 &=  \lambda^2_1\lambda^2_2  +  \lambda^2_1\lambda^2_3 +  \lambda^2_1\lambda^2_4+  \lambda^2_1\lambda^2_5 + \lambda^2_2\lambda^2_3 \\
&\quad + \lambda^2_2\lambda^2_4 + \lambda^2_2\lambda^2_5+ \lambda^2_3\lambda^2_4+ \lambda^2_3\lambda^2_5+ \lambda^2_4\lambda^2_5,\\
I_3 &= \lambda^2_1\lambda^2_2\lambda^2_3 
+  \lambda^2_1\lambda^2_2\lambda^2_4 
+  \lambda^2_1\lambda^2_2\lambda^2_5
+  \lambda^2_1\lambda^2_3\lambda^2_4 
+  \lambda^2_1\lambda^2_3\lambda^2_5
 +  \lambda^2_1\lambda^2_4\lambda^2_5 \\
  &\quad 
 +  \lambda^2_2\lambda^2_3\lambda^2_4 
 +  \lambda^2_2\lambda^2_3\lambda^2_5 
 +  \lambda^2_2\lambda^2_4\lambda^2_5 
 +  \lambda^2_3\lambda^2_4\lambda^2_5, \\
 I_4 &= \lambda^2_1\lambda^2_2\lambda^2_3\lambda^2_4
 + \lambda^2_1\lambda^2_2\lambda^2_3\lambda^2_5
 + \lambda^2_1\lambda^2_2\lambda^2_4\lambda^2_5
 + \lambda^2_1\lambda^2_3\lambda^2_4\lambda^2_5
 + \lambda^2_2\lambda^2_3\lambda^2_4\lambda^2_5,   \\
I_5 &= \lambda^2_1\lambda^2_2\lambda^2_3\lambda^2_4
\lambda^2_5.
\end{aligned}
\ee

The characteristic equation (\ref{2.4a}) for $\C$ becomes
\be \label{5.3a}
x^5-I_1x^4+I_2x^3-I_3x^2+I_4x-I_5=0,
\ee
the roots of which  are the five squared stretches.  Of course, there is no formula giving the roots of the quintic equation.

The invariants of $\U=\C^{1/2}$ are defined in terms of the principal stretches  by 
\be \label{5.4a}
\begin{aligned}
i_1 &= \lambda_1 + \lambda_2 + \lambda_3 + \lambda_4 + \lambda_5 ,\\
i_2 &=  \lambda_1\lambda_2  +  \lambda_1\lambda_3 +  \lambda_1\lambda_4+  \lambda_1\lambda_5 + \lambda_2\lambda_3 \\
&\quad + \lambda_2\lambda_4 + \lambda_2\lambda_5+ \lambda_3\lambda_4+ \lambda_3\lambda_5+ \lambda_4\lambda_5,\\
i_3 &= \lambda_1\lambda_2\lambda_3 
+  \lambda_1\lambda_2\lambda_4 
+  \lambda_1\lambda_2\lambda_5
+  \lambda_1\lambda_3\lambda_4 
+  \lambda_1\lambda_3\lambda_5
 +  \lambda_1\lambda_4\lambda_5 \\
  &\quad 
 +  \lambda_2\lambda_3\lambda_4 
 +  \lambda_2\lambda_3\lambda_5 
 +  \lambda_2\lambda_4\lambda_5 
 +  \lambda_3\lambda_4\lambda_5 \\
 i_4 &= \lambda_1\lambda_2\lambda_3\lambda_4
 + \lambda_1\lambda_2\lambda_3\lambda_5
 + \lambda_1\lambda_2\lambda_4\lambda_5
 + \lambda_1\lambda_3\lambda_4\lambda_5
 + \lambda_2\lambda_3\lambda_4\lambda_5,   \\
i_5 &= \lambda_1\lambda_2\lambda_3\lambda_4
\lambda_5.
\end{aligned}
\ee
The characteristic equation for $\U$ is
\[
x^5-i_1x^4+i_2x^3-i_3x^2+i_4x-i_5=0,
\]
the roots of which  are the five stretches.

We seek to determine the invariants  $i_\alpha$ of $\U$ in terms of the invariants  $I_\alpha$ of $\C$  and so derive
the following identities   in the same way that  (\ref{3.4a}) were derived:
\be \label{5.5a}
\begin{aligned}
i_1^2 &= I_1+2i_2,\\
i_2^2 &=  I_2+2i_1i_3-2i_4,\\
i_3^2 &= I_3+2i_2i_4-2i_1i_5,\\
i_4^2 &= I_4+2i_3i_5, \\
i_5^2 &=I_5.
\end{aligned}
\ee

We now seek to eliminate all but $i_1$ from equations  (\ref{5.5a}).  First use  (\ref{5.5a})$_{1,4}$ to eliminate $i_2$ and $i_3$ from  (\ref{5.5a})$_{2,3}$ in favour of $i_1$ and $i_4$.  Then  (\ref{5.5a})$_{2}$ becomes quadratic in $i_4$ but quartic in $i_1$ whereas (\ref{5.5a})$_{3}$ becomes quadratic in $i_1$ but quartic in $i_4$.  The quadratic (\ref{5.5a})$_{2}$ is solved for $i_4$ and the result used to eliminate $i_4$ from (\ref{5.5a})$_{3}$ in favour of $i_1$.
We find that $x=i_1$ is one root of the equation
\be\label{5.6a}
\pm 8 S\sqrt{R} = T,
\ee
where
\be \label{5.7a}
\begin{aligned}
R &= \sqrt{I_5}\,x^5-2I_1\sqrt{I_5}\, x^3 + 4I_4\,x^2 + (I_1^2-4I_2)\sqrt{I_5}\, x +4 I_5, \\
S &=  3x^5-2I_1x^3-(I_1^2-4I_2)x-8\sqrt{I_5} , \\
T &=   x^{10}-4I_1x^8+(6I_1^2-8I_2)x^6 +96\sqrt{I_5} \,x^5-4(I_1^3-4I_1I_2+16I_3)x^4\\ 
&\mbox{\quad}+ (I_1^4-8I_1^2I_2+16I_2^2+64I_4)x^2 
+32(I_1^2-4I_2)\sqrt{I_5} \,x +128I_5,
\end{aligned}
\ee
$R$ and $S$ being of the fifth degree in $x$ and $T$  of the tenth.  Squaring (\ref{5.6a}) leads~to
\be \label{5.8a}
T^2 - 64S^2R = 0,
\ee
a polynomial equation of the twentieth degree.  However, inspection of this polynomial reveals that the lowest power of $x$ occurring is $x^4$ so that the spurious quadruple root $x=0$ may be removed leaving in place of (\ref{5.8a}) a polynomial equation of the sixteenth degree, which we do not exhibit.

 We have seen that for $N=3$ and $N=4$ cubic and quartic equations suffice, whereas for $N=5$ we must go to an equation of degree 16.  The reason for this is now made clear.

As before, we note that $\sqrt{I_5} = i_5$ is invariant under sign change of any \emph{two} of the $\lambda_i$, so that equations   (\ref{5.6a})--(\ref{5.8a}) are similarly invariant.   Therefore the 16 non-zero roots of (\ref{5.8a}) must have the same invariance.  They are:
\be \label{5.9a}
\begin{aligned}
&\lambda_1+\lambda_2+\lambda_3+\lambda_4+\lambda_5\\
&\lambda_1+\lambda_2+\lambda_3-\lambda_4-\lambda_5\\
&\dots \\
&\lambda_1-\lambda_2-\lambda_3-\lambda_4-\lambda_5\\
&\dots
\end{aligned}
\ee
only the first of which, the largest, gives $i_1$.  There are a further ten roots with two minus signs and then a further five roots with four minus signs, making 16 in all.  

Numerical evidence, for example with $\lambda_1 = 5, \lambda_2 = 4, \lambda_3 = 3, \lambda_4 = 2, \lambda_5 = 1,$ bears out these conclusions.     

\subsection{$\U$ and $\U^{-1}$ expressed in powers of $\C$ and the invariants of $\U$} 

For $N=5$ we need to define
\be\label{5.10a}
\nu_5 = i_1i_2i_3i_4 + 2i_1i_4i_5+i_2i_3i_5 - i_1^2i_4^2  - i_1i_2^2i_5 - i_3^2i_4  - i_5^2.
\ee
We can show that $(\lambda_i+\lambda_j)$ for $i \neq j$ is a factor of $\nu_5$.  There are 10 such factors and we have
\be\label{5.11a}
\nu_5 = \prod_{i,\,j=1,\;i<j}^5 (\lambda_i+\lambda_j).
\ee
We take the Cayley-Hamilton theorem for $\U$ and multiply it successively by $\U$, $\U^2$ and $\U^3$.  From the resulting four equations we eliminate $\U^7$, $\U^5$ and $\U^3$ in favour of $\U$ to obtain
\be\label{5.12a}
\nu_5\U = p_0\I+p_1\C-p_2\C^2+p_3\C^3-p_4\C^4,
\ee 
where the coefficients are given in terms of the invariants $i_1\dots i_5$ by
\be\label{5.13a}
\begin{aligned}
 p_0 &= ( i_1i_2i_3 + i_1i_5 - i_1^2i_4 - i_3^2 ) i_5\\
 p_1 &= i_1^2i_2i_5 + i_1i_2i_3^2   + i_1i_4^2 + i_2i_3i_4
            - i_1^2i_3i_4 - i_1i_2^2i_4 - i_3^3 - i_4i_5 \\
 p_2 &= i_1^3i_4  + i_1i_2^3 + 2i_1i_3^2 + i_2i_5
            -2i_1^2i_2i_3  - i_1^2i_5 - i_2^2i_3 - i_3i_4 \\
 p_3 &= i_1^3i_2 + i_1i_4 + 2i_2i_3 - i_1^2i_3 - 2i_1i_2^2 - i_5 \\
 p_4 &= i_1i_2 - i_3.
\end{aligned}
\ee
On multiplying (\ref{5.12a}) by $\C^{-1}$ and arguing as before we obtain
\be\label{5.14a}
\begin{aligned}
\nu_5\U^{-1} &= \left(p_1+p_0\frac{I_4}{I_5} \right) \I
 - \left(p_2+p_0\frac{I_3}{I_5} \right) \C
 + \left(p_3+p_0\frac{I_2}{I_5} \right) \C^2  \\
&\quad - \left(p_4+p_0\frac{I_1}{I_5} \right) \C^3
 + \frac{p_0}{I_5}\C^4.
\end{aligned}
\ee
Equations (\ref{5.12a}) and (\ref{5.14a})  express $\U$ and $\U^{-1}$, respectively,  in terms of powers of $\C$, though it does not seem possible to express the coefficients in terms of the invariants of $\C$ and so they are left in terms of the invariants of~$\U$, see (\ref{5.13a}).


\section{An equation for $i_1$ in terms of the invariants of $\C$ for $N=6$}  
\label{sec:6}
\setcounter{equation}{0}

For $N=6$, the six principal invariants $I_\alpha$ of $\C$ and $i_\alpha$ of $\U$  are defined similarly to  (\ref{5.2a}) and (\ref{5.4a}), respectively, for $N=5$.  For example,
\be\label{6.1a}
\begin{aligned}
i_1 &= \lambda_1+\lambda_2+\lambda_3+\lambda_4+\lambda_5+\lambda_6, \\
i_6 &=  \lambda_1\lambda_2\lambda_3\lambda_4\lambda_5\lambda_6. 
\end{aligned}
\ee

The characteristic equation (\ref{2.4a}) for $\C$, and that for $\U$, become, respectively,
\be \label{6.2a}  
\begin{aligned}
 x^6-I_1x^5+I_2x^4-I_3x^3+I_4x^2-I_5x+I_6 &=0, \\
 x^6-i_1x^5+i_2x^4-i_3x^3+i_4x^2-i_5x+i_6 &=0.  
 \end{aligned}
 \ee
Of course, there is no formula giving the roots of these sextic equations.

We seek to determine, as far as possible, the invariants $i_\alpha$ of $\U$ in terms of the invariants $I_\alpha$ of $\C$  and so derive
the following identities  from the characteristic equations (\ref{6.2a}) in the same way that  (\ref{3.4a}) were derived:
\be \label{6.3a}
\begin{aligned}
i_1^2 &= I_1+2i_2,\\
i_2^2 &=  I_2+2i_1i_3-2i_4,\\
i_3^2 &= I_3+2i_2i_4-2i_1i_5+2i_6,\\
i_4^2 &= I_4+2i_3i_5-2i_2i_6, \\
i_5^2 &= I_5+2i_4i_6, \\
i_6^2 &=I_6.
\end{aligned}
\ee
We wish to eliminate $i_2,i_3,i_4,i_5$ from equations (\ref{6.3a}) in favour of $i_1$, $i_6$ and the~$I_\alpha$.   We begin by using (\ref{6.3a})$_2$ to eliminate $i_3$ from the other equations. We then use the resulting equation (\ref{6.3a})$_3$ to eliminate $i_5$ from the remaining equations to obtain the following equations, a cubic in $i_4$ and a quartic in $i_4$,
\be\label{6.4a}
\begin{aligned}
i_4^3+a_2 i_4^2 +a_4 i_4 +a_6 &=0  \\
i_4^4 +b_2i_4^3+b_4i_4^2+b_6i_4+b_8 &=0
\end{aligned}
\ee
with coefficients defined by
\be\label{6.5a}
\begin{aligned}
a_2 &= \tfrac32 (i_2^2-I_2) - 2i_1^2i_2  + i_1^4   \\
a_4 &= \tfrac34 (i_2^2-I_2)^2 - i_1^2i_2(i_2^2-I_2) - i_1^2 (I_3+2i_6) \\
a_6 &= \tfrac18 (i_2^2-I_2)^3 - \tfrac12 i_2^2(i_2^2-I_2) (I_3+2i_6) + i_1^4(2i_2i_6-I_4) \\
b_2 &= 2(i_2^2-I_2-2i_1^2i_2) \\
b_4 &=  (i_2^2-I_2-2i_1^2i_2)^2 + \tfrac12 (i_2^2-I_2)^2  - 2i_1^2(I_3+2i_6)   \\
b_6 &= \tfrac12  (i_2^2-I_2-2i_1^2i_2) [ (i_2^2-I_2)^2 - 4i_1^2(I_3+2i_6) ] - 8i_1^6i_6   \\
b_8 &=  \tfrac{1}{16}  [ (i_2^2-I_2)^2 - 4i_1^2(I_3+2i_6) ]^2 - 4 i_1^6I_5.
\end{aligned}
\ee
These coefficients depend on $i_1$ through $w=i_1^2$ and $i_2=(w-I_1)/2$, see (\ref{6.3a})$_1$, and each  is a polynomial in $w$ with the subscript label on the left equal to the degree of the polynomial  on the right.   These coefficients depend only on the $I_\alpha$ and~$i_6$.

We now wish to eliminate $i_4$ from (\ref{6.4a}) to obtain an equation for $w$.  We therefore multiply (\ref{6.4a})$_1$ by $i_4$ and subtract it from (\ref{6.4a})$_2$ to obtain another cubic equation in $i_4$.  We proceed like this, subtracting multiples of one equation from another, until we arrive at
\be\label{6.6a}
\begin{aligned}
c_4 i_4^2 +c_6 i_4 +c_8 &=0  \\
d_{12}i_4+d_{14} &=0,
\end{aligned}
\ee
in which we have defined
\be\label{6.7a}
\begin{aligned}
c_4  &=  (b_2-a_2)a_2 - ( b_4-a_4) \\
c_6  &=  (b_2-a_2)a_4 - ( b_6-a_6)  \\
c_8 &= (b_2-a_2)a_6 -  b_8  \\
d_{12} &= (a_4c_4-c_8)c_4 - (a_2c_4-c_6)c_6  \\
d_{14} & = a_6c_4^2 - (a_2c_4-c_6)c_8
\end{aligned}
\ee
the right hand sides being polynomials in $w$ of degree indicated by the subscript on the left.

We eliminate  $i_4$ by substituting for it  from (\ref{6.6a})$_2$ into (\ref{6.6a})$_1$ giving
\be\label{6.8a}
e_{32}^{\phantom{32}} \equiv -2^{30}\left( c_8d_{12}^2 - c_6d_{12}d_{14}  + c_4d_{14}^2 \right) = 0,
\ee
a polynomial of degree 32 in $w$.   With the aid of a computer algebra package, we choose the numerical prefactor $-2^{30}$  so that the polynomial shall have leading term $121w^{32}$ and the coefficients of the lower powers of $w$ turn out to be linear combinations of integer multiples of products of the $I_\alpha$ and $i_6$. 
Equation (\ref{6.8a}) is the equivalent for $N=6$ of the 20 degree polynomial  (\ref{5.8a}) for $N=5$.  Using the computer algebra package (\ref{6.8a}) has been evaluated for general values of the $I_\alpha$ and $i_6$ and for all such values the lowest power of $w$ occurring is $w^{10}$ and so (\ref{6.8a}) has 10 spurious zeros, just as (\ref{5.8a}) has four.

Equations (\ref{6.5a})--(\ref{6.8a}) depend on $i_6$ and so are invariant under \emph{two} sign changes among the $\lambda_i$.  Therefore the roots of (\ref{6.8a}) must be similarly invariant.  Consider possible values of $i_1$:
\be \label{6.9a}
\begin{aligned}
&\lambda_1+\lambda_2+\lambda_3+\lambda_4+\lambda_5+\lambda_6\\
&\lambda_1+\lambda_2+\lambda_3+\lambda_4-\lambda_5-\lambda_6\\
&\dots \\
&-\lambda_1-\lambda_2+\lambda_3+\lambda_4+\lambda_5+\lambda_6\\
\hline
&+\lambda_1+\lambda_2-\lambda_3-\lambda_4-\lambda_5-\lambda_6\\
&\dots \\
&-\lambda_1-\lambda_2-\lambda_3-\lambda_4-\lambda_5-\lambda_6
\end{aligned}
\ee
only the first of which, the largest, gives $i_1$.  The next 15 (above the horizontal line) give all possibilities with two minus signs, the 15 immediately below the horizontal line give all possibilities with four minus signs and the final line is the only possibility with  six minus signs.  This makes 32 possibilities in total but each of the last 16 is the negative of one of the first 16.  This explains why $w=i_1^2$ occurs in equations (\ref{6.5a})--(\ref{6.8a}) rather than just $i_1$.  Thus 16 of the roots of (\ref{6.8a}) are the squares of quantities to be found in the table (\ref{6.9a}), which together with the 10 spurious zero roots implies the existence of six further spurious roots. 

The six further spurious roots are given by the three roots of the cubic equation
\be\label{6.10a}   11w^3-7I_1w^2+(5I_1^2-12I_2)w - (I_1^3-4I_1I_2+8I_3+16i_6) = 0,
\ee
each repeated to make the total of six.  The quantity $w$ is quadratic in the $\lambda_i$ and so the coefficient of $w^2$ must also be a quadratic in the $\lambda_i$ and so must be a multiple of $I_1$.  The coefficient of $w$ is quartic in the $\lambda_i$ and so must be a linear combination of $I_1^2$ and $I_2$, and similarly for the constant term.  The integer multipliers are determined using the computer algebra package.

 In the examples below, each of the 16 linear factors before the $\times$ sign  corresponds to the square of an entry in the first 16 rows of equation  (\ref{6.8a}), the first, and largest, factor corresponding to $i_1^2$.  The 16 factors after the $\times$ sign are the 10 spurious zero factors and the six further spurious factors.
\begin{enumerate}
\item 
For equal principal stretches $\{1,1,1,1,1,1\}$, (\ref{6.8a}) has the factors
\[ (w-6^2) (w-2^2)^{15} \times w^{10} (w-2^2)^2 (11w^2+2w+8)^2.\]
The first factor corresponds to the correct value of $i_1=6$.  The next 15 factors correspond to $i_1=2$, obtained by taking all possibilities when any two of the 1's are replaced by $-1$.  The factor $(w-2^2)$ also appears as a spurious factor after the $\times$ sign.  The final quadratic factor has a complex conjugate pair of~zeros.
\item 
For principal stretches $\{1,2,3,4,5,6\}$, (\ref{6.8a}) has the factors
\[ \begin{aligned} (w-21^2) (w-15^2)(w-13^2)(w-11^2)^2(w-9^2)^2(w-7^2)^3 ((w-5^2)^2 \\ (w-3^2)^2(w-1)^2  
\times w^{10}(w-7^2)^2 (11w^2-98w+567)^2. 
\end{aligned} \]
The first factor gives correctly $i_1=21$.  The next 15 factors are obtained by taking all possibilities in (\ref{6.8a}) when any two of the $\lambda_i$ have their sign reversed.   The  factor $(w-7^2)$ also appears as a spurious factor after the $\times$ sign.  Once again, the final quadratic factor has a complex conjugate pair of zeros.
\item 
For principal stretches $\{1,2,3,5,6,7\}$,  (\ref{6.8a}) has the factors
\[ \begin{aligned}   (w-24^2) (w-18^2) (w-16^2) (w-14^2)  (w-12^2) (w-10^2)^2 
(w-8^2)^3 \\
  (w-6^2)^2 (w-4^2)(w-2^2)^2 w \times w^{10}  (w-8^2)^2 (11w^2-164w+648)^2.
 \end{aligned}  \]
 Because $1+2+3-5+6-7=0$ the factor $w$ occurs also before the $\times$ sign.
\item 
For principal stretches $\{1,2,4,5,7,8\}$,  (\ref{6.8a}) has the factors
\[ \begin{aligned}
 (w-27^2)  (w-21^2) (w-17^2) (w-15^2)^2   (w-13^2)  (w-11^2)  (w-9^2)^3 \\  (w-7^2)  
(w-5^2) (w-3^2)^3 (w-1) \times w^{10} (w-3^2)^2  (w-9^2)^2  (11w-123)^2.
 \end{aligned} \]
 The spurious factors are now all real and linear.
\item 
For principal stretches $\{1,1,1,2,5,7\}$,  (\ref{6.8a}) has the factors
\[ \begin{aligned}
(w-17^2)   (w-13^2)^3  (w-11^2)^3  (w-7^2)  
(w-5^2)^3   (w-3^2) (w-1)^4  \\  \mbox{}  \times w^{10} 
(11w^3-567w^2+11709w-40553)^2.
 \end{aligned} \]
 The final cubic factor has one  irrational real zero and a complex conjugate~pair.
\end{enumerate}
In all cases we can verify the formula (\ref{6.10a}) for the spurious roots.

It is possible to divide the 32 degree polynomial (\ref{6.8a}) by the 16 degree polynomial whose roots are the the 10 spurious zeros and the six spurious zeros coming from the square of the cubic (\ref{6.10a}) to get the following equation for $w$:
\be\label{6.11a}
\begin{aligned}
w^{16} &-16I_1w^{15} +(120I_1^2-32I_2)w^{14} \\
           & - (560I_1^3-448I_1I_2+256I_3+3840i_6)w^{13} + \cdots = 0.
\end{aligned}
\ee
This equation is too long to exhibit in full.  Note that the coefficients here have the same dimensional form as those in (\ref{6.10a}), as expected.

\subsection{$\U$ and $\U^{-1}$ expressed in powers of $\C$ and the invariants of $\U$} 

For $N=6$ we define
\be\label{6.12a}
\begin{aligned}
\nu_6 &= i_1i_2i_3i_4i_5 +  i_1^2i_3i_4i_6 +  2i_1^2i_2i_5i_6 + 2i_1i_4i_5^2 + i_3^3i_6 + i_2i_3i_5^2  \\
          &\;\;   - i_1^3i_6^2 -  i_1i_2i_3^2i_6 -3 i_1i_3i_5 i_6 - i_1^2i_4^2i_5 - i_1i_2^2i_5^2- i_3^2i_4i_5 - i_5^3
\end{aligned}
\ee
which has the factor $(\lambda_i+\lambda_j)$ for all $i \neq j$.  There are 15 such factors and we have
\be\label{6.13a}
\nu_6 = \prod_{i,\,j=1,\;i<j}^6 (\lambda_i+\lambda_j).
\ee
As before, we take multiples of the Cayley-Hamilton theorem for $\U$ to obtain
\be\label{6.14a}
\nu_6\U = p_0\I+p_1\C - p_2\C^2+p_3\C^3 - p_4\C^4+p_5\C^5,
\ee 
where the coefficients are given in terms of the invariants $i_1\dots i_6$ by
\be\label{6.15a}
\begin{aligned}
 p_0 &= ( i_1i_2i_3i_4 + 2i_1i_4i_5 + i_2i_3i_5 + i_1^2i_2i_6 -i_1i_2^2i_5 - i_1i_3i_6 - i_1^2i_4^2 - i_3^2i_4 - i_5^2  ) i_6\\
 p_1 &=  - i_1^3i_5i_6+i_1^2i_2i_5^2+i_1^2i_3i_4i_5 - i_1^2i_4^3+i_1^2i_6^2 +i_1i_2^2i_3i_6 -i_1i_2^2i_4i_5 -i_1i_2i_3^2i_5  \\
        &\quad +i_1i_2i_3i_4^2 -i_1i_3i_4i_6 -2i_1i_3i_5^2 +2i_1i_4^2i_5 -i_2i_3^2i_6+i_2i_3i_4i_5 +i_3^3i_5 -i_3^2i_4^2    \\
        &\quad   +i_3i_5i_6 -i_4i_5^2\\
 p_2 &=   i_1^3i_3i_6 -i_1^3i_4i_5 -i_1^2i_2^2i_6 +2i_1^2i_2i_4^2 -i_1^2i_3^2i_4 +i_1^2i_5^2 +i_1i_2^3i_5 -2i_1i_2^2i_3i_4   \\         
       &\quad  +i_1i_2i_3^3   -2i_1i_2i_4i_5 +i_1i_3^2i_5 -i_1i_5i_6 -i_2^2i_3i_5 +2i_2i_3^2i_4 +i_2i_5^2  \\ 
       &\quad  -i_3^4 +i_3^2i_6 -i_3i_4i_5   \\
 p_3 &=  -i_1^4i_6 +i_1^3i_2i_5 +2i_1^3i_3i_4 -i_1^2i_2^2i_4 -2i_1^2i_2i_3^2 +2i_1^2i_2i_6 -3i_1^2i_3i_5 -2i_1^2i_4^2  \\
        &\quad + i_1i_2^3i_3  -i_1i_2^2i_5 +2i_1i_2i_3i_4 +2i_1i_3^3 -i_1i_3i_6 +3i_1i_4i_5 -i_2^2i_3^2 +2i_2i_3i_5  \\
        &\quad  -2 i_3^2i_4 -i_5^2  \\
 p_4 &=  -i_1^4i_4 +i_1^3i_2i_3 +i_1^3i_5 +2i_1^2i_2i_4 -i_1^2i_3^2 -i_1^2i_6 -2i_1i_2^2i_3 -i_1i_2i_5 +2i_2i_3^2-i_3i_5   \\
 p_5 &= i_1i_2i_3 +i_1i_5 -i_1^2i_4 -i_3^2.
\end{aligned}
\ee
On multiplying (\ref{6.14a}) by $\C^{-1}$ and arguing as before we obtain
\be\label{6.16a}
\begin{aligned}
\nu_6\U^{-1} &= \left(p_1+p_0\frac{I_5}{I_6} \right) \I
 - \left(p_2+p_0\frac{I_4}{I_6} \right) \C
 + \left(p_3+p_0\frac{I_3}{I_6} \right) \C^2  \\
&\quad - \left(p_4+p_0\frac{I_2}{I_6} \right) \C^3
+ \left(p_5+p_0\frac{I_1}{I_6} \right) \C^4
 - \frac{p_0}{I_6}\C^5.
\end{aligned}
\ee
Equations (\ref{6.14a}) and (\ref{6.16a})  express $\U$ and $\U^{-1}$, respectively,  in terms of powers of $\C$, though it does not seem possible to express the coefficients  (\ref{6.15a}) in terms of the invariants of $\C$ and so they are left in terms of the invariants of~$\U$.


\section{Discussion} 
\label{sec:7}
\setcounter{equation}{0}

If we may assume for odd $N=2M+1$ that we can obtain a polynomial equation in $x$ for $i_1$ with coefficients dependent only on the $I_\alpha$ and $i_N = \lambda_1\lambda_2\ldots \lambda_N$ then as before the set of  roots of this equation is invariant under sign change of any $2,4,\ldots, 2M$ of the  $\lambda_1, \lambda_2,\ldots,  \lambda_N$.  The number of such roots is
\[ \sum_{m=0}^{M} {2M+1 \choose 2m }  = 2^{2M},  \]
see equation (\ref{5.9a}), for example, where $M=2$.
Therefore the polynomial equation in $x$ for $i_1$ must be of  degree $4^M$, possibly higher if there are spurious roots.

For even $N=2M+2$ we would similarly expect invariance under sign change of $2,4,\ldots, 2M+2$ of the  $\lambda_1, \lambda_2,\ldots,  \lambda_N$.  The number of such roots is
\[ \sum_{m=0}^{M+1} {2M+2 \choose 2m }  = 2^{2M+1},  \]
see equation (\ref{6.8a}), for example, where $M=2$.
Therefore the polynomial equation in $x$ for $i_1$ must be of degree at least $2^{2M+1}$.  However, as for $N=6$,  each of these roots occurs with its negative and so we actually have a polynomial of degree $4^M$ in $w=i_1^2$, higher if there are spurious~roots.

For each unit increase in $M$ we can see that the degree of the polynomial for $i_1$ or $i_1^2$ is quadrupled.  Thus, we have quartic equations for $3\times3$ and $4\times4$ tensors,  equations of degree 16 for $5\times5$ and $6\times6$ tensors and  of at least degree 64 for $7\times7$ and $8\times8$ tensors.

\end{document}